\documentclass[english,12pt]{article}
          \usepackage{babel}
          \usepackage[T1]{fontenc}
          \usepackage[latin2]{inputenc}
          \usepackage{pslatex}
\begin{document}
\begin{flushright}
\textit{ACTA ASTRONOMICA}\\
Vol.0 (2015) pp.0-00
\end{flushright}
\bigskip

\begin{center}
{\bf Hot spot in eclipsing dwarf nova IY Ursae Majoris\\ during quiescence and normal outburst}\\
\vspace{0.8cm}
{K. ~B~\k{a}~k~o~w~s~k~a$^{1,2}$ \& A. ~O~l~e~c~h$^1$}\\
\vspace{0.5cm}
\begin{small}
{$^1$ Nicolaus Copernicus Astronomical Center,
Polish Academy of Sciences,\\
ul.~Bartycka~18, 00-716~Warszawa, Poland\\}
{$^2$ Fulbright Visiting Scholar, The Ohio State University\\  Dept. of Astronomy, 140 W. 18th Ave, Columbus, OH 43210, USA\\}
\end{small}
{\tt e-mail: bakowska@camk.edu.pl}\\
~\\
\end{center}

\begin{abstract}  

We present the analysis of hot spot brightness in light curves of the eclipsing dwarf nova IY Ursae Majoris during its normal outburst in March 2013 and in quiescence in April 2012 and in October 2015. Examination of four reconstructed light curves of the hot spot eclipses showed directly that the brightness of the hot spot changed significantly only during the outburst. The brightness of the hot spot, before and after the outburst, was on the same level. Hereby, based on the behaviour of the hot spot, IY UMa during its normal outburst follows the disk-instability model.  

\noindent {\bf Key words:} \textit{Stars: individual: IY UMa - binaries: 
close - novae, cataclysmic variables}

\end{abstract}

\section{Introduction}

Among close binaries there is a group of cataclysmic variable stars (CVs) consisting of a white dwarf (the primary) that accretes matter from a secondary star via Roche lobe overflow. The secondary component usually resembles a low-mass main-sequence star. The variety of observed features in the light curves of CVs leads to a sophisticated taxonomy (Warner 1995). In CVs with a weak magnetic field ($<10^6$ G) an accretion disk is formed around the primary. The hot spot (or bright spot) is the region where the accreted matter collides with the edge of the disk. 

Based on the photometric behaviour, among CVs there is a group called dwarf novae (DNs) of a SU UMa type. This class of objects has short orbital periods ($P_{orb} < 2.5$ h). In the light curves of SU UMa type stars there are present regular sudden increases of brightness called outbursts or superoutbursts. Outbursts are more frequent then superoutbursts. Also, observed brightness of SU UMa stars during superoutbursts is about one magnitude higher than during outbursts. Additionally, in the light curves of SU UMa type DNs during superoutbursts there are characteristic periodic light oscillatons called superhumps (more in Hellier 2001).

IY UMa was discovered by Takamizawa (1998, vsnet-obs circulation 18078). This variable was classified as a new deeply eclipsing SU UMa type DN after its January 2000 superoutburst with clear superhumps manifestation (Uemura et al. 2000, Patterson et al. 2000, Stanishev et al. 2001). Steeghs et al. (2003) conducted high speed photometry of IY UMa during quiescence and based on analyzed eclipses the most up-to-date orbital period of $P_{orb}=0.07390897(5)$ days was derived. Moreover, in the light curves of eclipses of IY UMa they detected a curious short-lived rise of brightness  which occurs between the end of white dwarf ingress and hot spot ingress. Data presented by Steeghs et al. (2003) were reanalyzed by Smak (2003) who postulated stream overflow for this eclipsing DN. About normal outbursts in IY UMa only two publications have been published so far. Three normal outbursts with duration around $2-3$ days, in April 2000, December 2000 and March 2001 were reported by Kato et al. (2001).  Also, Rolfe et al. (2005) observed one normal outburst of IY UMa during two nights in January 2001. Detection of these eruptions indicates that IY UMa goes into normal outburst several times per year. Nonetheless, the hot spot brightness of IY UMa has not been investigated before, during and after normal outbursts, hence our motivation for this analysis. 

This paper is arranged as follows: in Section 2 there is information about observations, data reduction and photometry analysis. Section 3 presents global photometric behaviour of IY UMa. Later, in Section 4 we show the result of the decomposition analysis. Discussion is presented in Section 5, followed by our conclusions.

\section{Observations and data reduction}

Observations of IY UMa presented in this work were gathered during 6 nights between 2012 April 17 and 2015 October 15. In this campaign we observed the variable in quiescence (2 nights - 2012 April 17/18 and 23/24), during one normal outburst (1 night - 2013 March 05/06) and again at minimum brightness (3 nights - 2013 March 06/07, 2015 October 14/15 and 15/16).

Table 1 presents a full journal of our CCD observations of IY UMa.  The object was monitored during 14.16 hours. We obtained 1069 useful exposures. 

All data from 2012 and 2013 were collected at the Ostrowik station of the Warsaw University Observatory. IY UMa was observed on 60-cm Cassegrain telescope equipped with a Tektronics TK512CB back-illuminated CCD camera. The scale of the camera was 0.76 arcsec/pixel providing a 6'.5 x 6'.5 field of view. The description of the camera and the telescope was presented by Udalski and Pych (1992). To obtain the shortest possible exposure times the object was observed in "white light". The exposure times varied from 80 to 120 seconds depending of the brightness of the star, the seeing and the sky transparency.

Observations of IY UMa from 2015 were obtained at the MDM Observatory with the 1.3 meter McGraw Hill telescope located on Kitt Peak Mountain in Arizona.  Data were collected with a thinned, backside-illuminated Ikon DU-937N CCD camera built by Andor Technology PLC. The active area was 512 x 512 pixels, each 13 $\mu$m square. Exposure times were between 10 and 15 seconds and the dead time between exposures was only 11.92 ms. The star was monitored in a Schott GG420 filter. 

The IRAF\footnote{IRAF is distributed by
the National Optical Astronomy Observatory, which is operated by the
Association of Universities for Research in Astronomy, Inc., under a
cooperative agreement with the National Science Foundation.} package was used for the standard correction for bias, flat-field and dark frames. DAOPHOTII package (Stetson 1987) was used for profile photometry.  

The relative unfiltered magnitude of IY UMa was obtained by taking the difference between the magnitude of the object and the mean magnitude of two comparison stars. In Fig.1 the map of the observed region is presented where IY UMa is marked as V1 and the two comparison stars are marked as C1 and C2, respectively. The typical
accuracy of our measurements varied between 0.029 mag and 0.174 mag for data set from Poland and from 0.013 mag to 0.043 mag for observations from the U.S.A., depending on the weather conditions and the brightness of the star. The median value of the photometric errors was 0.054 mag and 0.027 mag for Ostrowik and MDM Observatory, respectively.
	
\begin{table*}[!ht]
 \centering
 \begin{small}
  \caption{Journal of our CCD observations of IY UMa.}
\smallskip
  \begin{tabular}{@{}|c|c|c|c|c|c|@{}}
  \hline
Date   & Time of start   & Length     & Number of  &  Telescope & Observer   \\
       & 2450000+ [HJD]  & of run [h] & frames     
&         &   \\
\hline
2012 Apr 17 & 6035.29211 & 2.65 & 88 & Ostrowik (Poland) & K. B\k{a}kowska, A. Olech\\
2012 Apr 23 & 6041.31852 & 0.84 & 29 & Ostrowik (Poland) & K. B\k{a}kowska\\
2013 Mar 05 & 6356.45747 & 4.06 & 160 & Ostrowik (Poland) & K. B\k{a}kowska\\
2013 Mar 06 & 6357.44345 & 4.36 & 160 & Ostrowik (Poland) & K. B\k{a}kowska\\
2015 Oct 14 & 7310.95955 & 1.00 & 334 & MDM Obs. (USA) & K. B\k{a}kowska\\
2015 Oct 15  & 7311.95956 & 1.25 & 298 &  MDM Obs. (USA) & K. B\k{a}kowska\\
\hline
\end{tabular}
 \end{small}
\label{tab0}
\end{table*}   

\vspace*{10.0cm}
\includegraphics{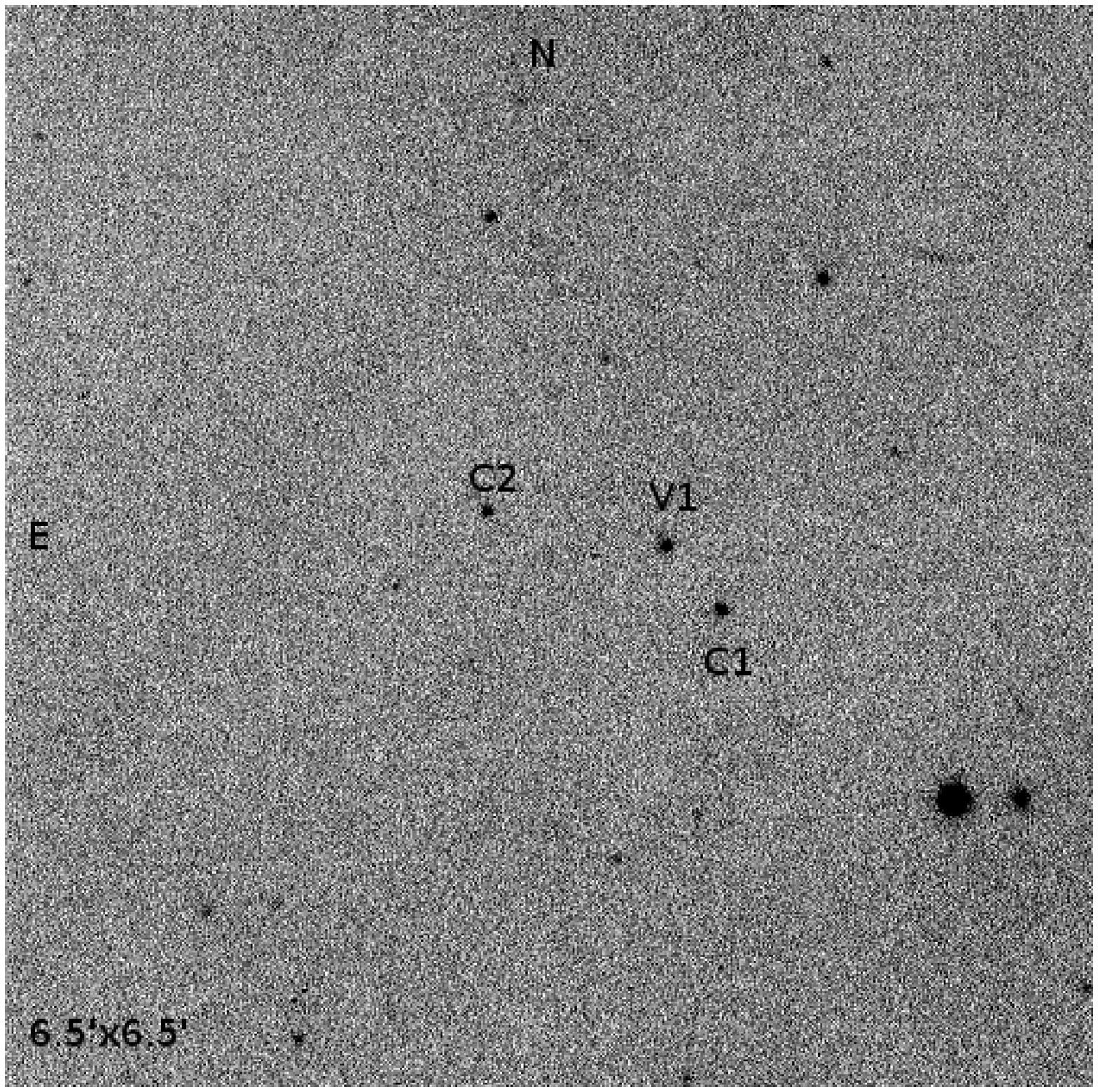}

   \begin{figure}[!ht]
      \caption {Finding chart of IY UMa, marked as V1. The positions of both comparison stars are also shown as C1 and C2, respectively. The field of view is about 6.5'$\times$6.5'. North is up, east is left.}
   \end{figure}

\section{Light curves}

Fig.2 shows the global photometric behaviour of IY UMa between 2012 April 17/18 (top panel) and 2015 October 15/16 (bottom panel). On 2013 March 05/06 we detected the object during the last night of the normal outburst and the rise of brightness of IY UMa during its eruption was about $\sim0.95$ mag comparing to the quiescent level.

We obtained light curves with full coverage of eclipses of IY UMa during four nights and only these data were used for the analysis of the hot spot manifestation. During three runs: 17/18 April 2012, 05/06 March 2013 and 06/07 March 2013, each time we monitored the variable during two subsequent eclipses. Hence, for these nights data were phased to obtain a better coverage of successive stages of eclipses. In Fig.3 we present eclipses of IY UMa chosen for further analysis. 

The data for the first eclipse (top left panel) was obtained during the night when IY UMa was in quiescence. This enabled us to derive the hot spot brightness before the outburst. Observations from 2012 March 05/06  (top right panel) were collected during the last night of the outburst of the eclipsing DN. The next eclipse (bottom left panel) was recorded on the first night of IY UMa in quiescence after the March 2012 outburst. The last eclipse (bottom right panel) was also from the quiescent stage of the variable. Worth noting is that in the light curve of this eclipse we detected the rise of brightness between the end of white dwarf ingress and hotspot ingress reported earlier by Steeghs et al. (2003).

\vspace*{15.0cm}
\includegraphics{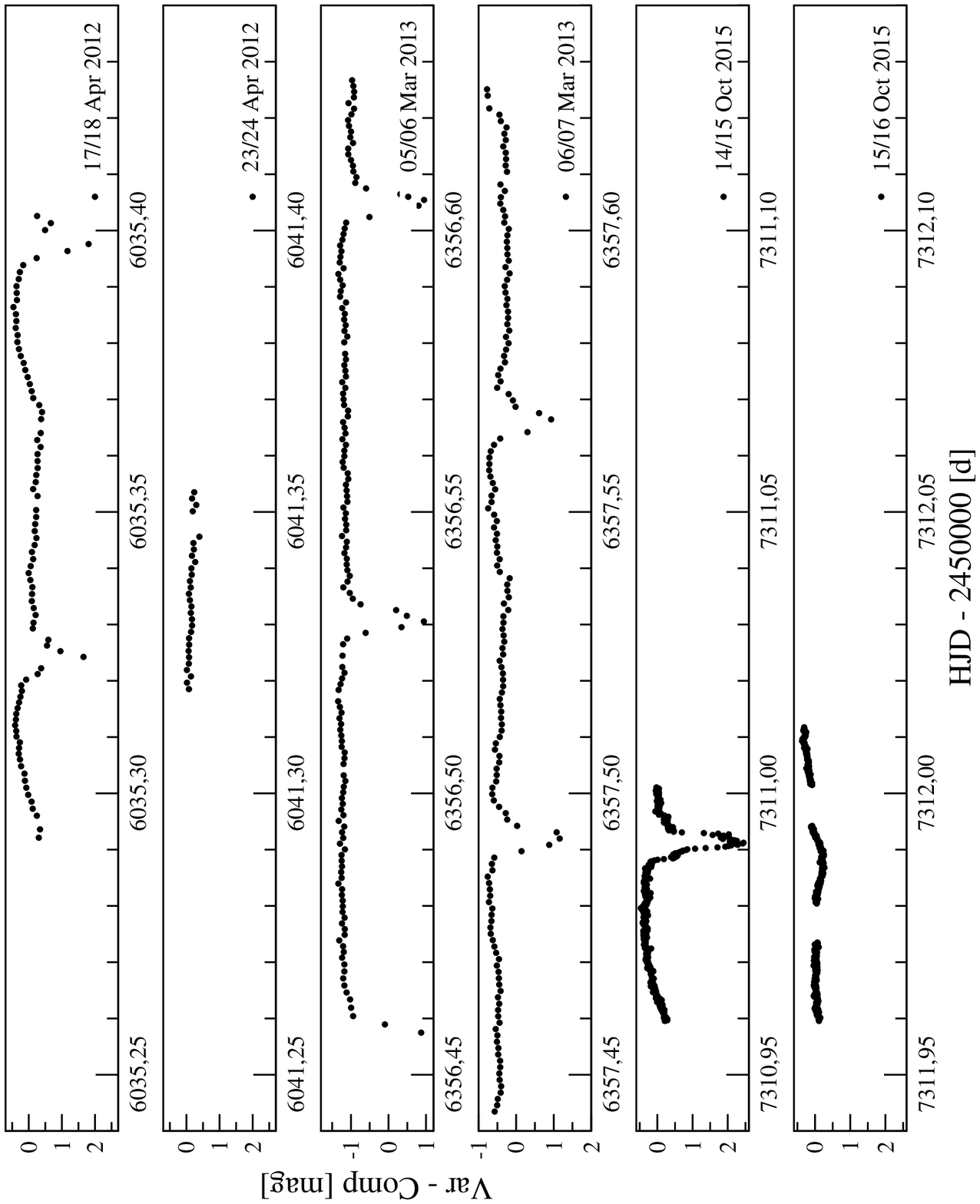}

   \begin{figure}[!ht]
      \caption {The global photometric behaviour of IY UMa. The first and the second panels (from the top) present two nights of IY UMa in quiescent state. The next panel shows one night of the normal outburst (2013 March 05/06). In the last three bottom panels the light curves from the quiscent state of IY UMa are presented.}
   \end{figure}  

\newpage
\vspace*{15.0cm}
\includegraphics{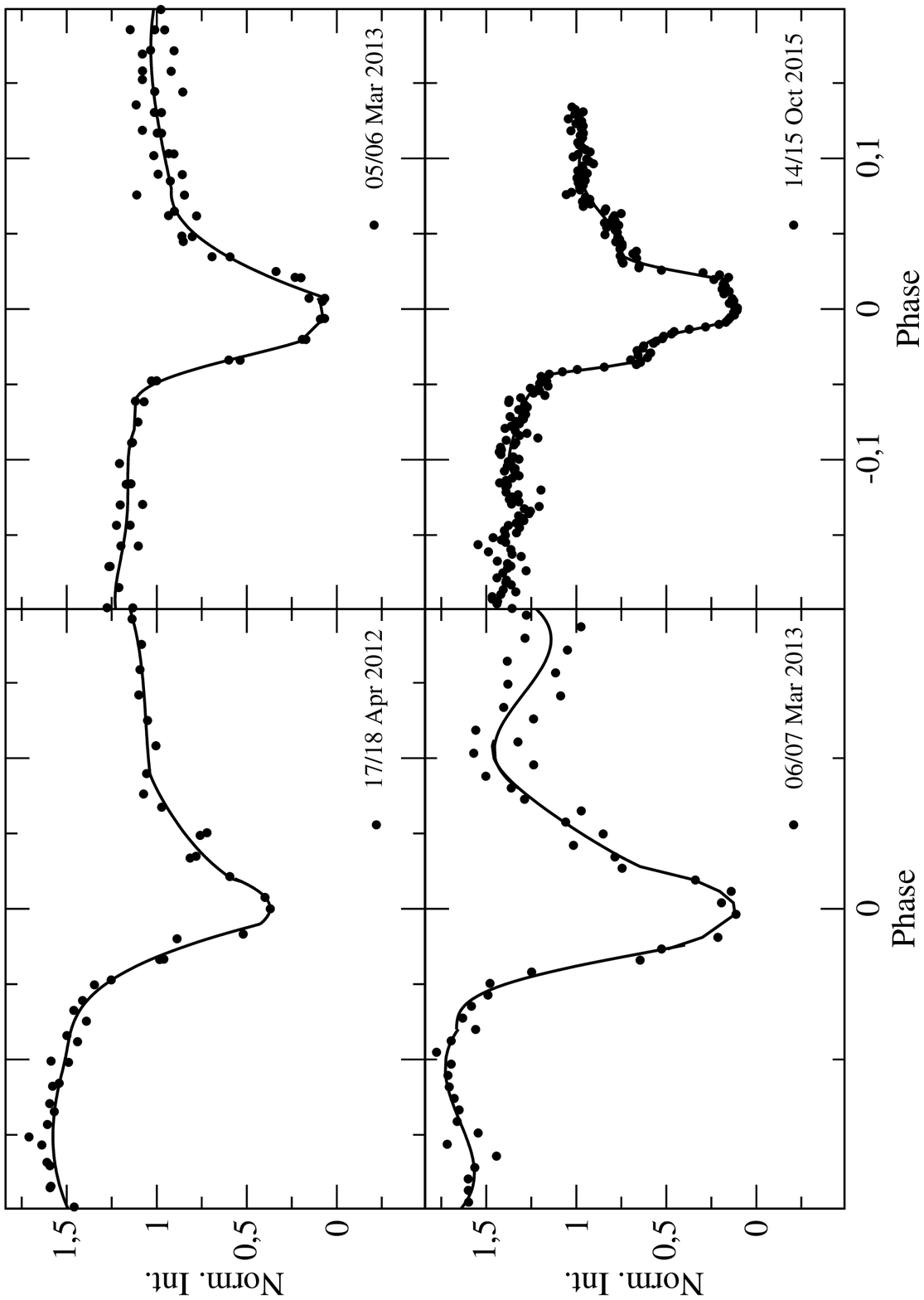}

   \begin{figure}[!ht]
      \caption {Eclipses chosen for decomposition analysis are presented. Observational data are displayed as black circles. Black lines represent synthetic light curves derived from the higher order polynomial fits. Those synthetic light curves were used for further analysis where separate light curves from the hot spot and the disk were obtained.}
   \end{figure}

\section{Results}

Smak (1994a) invented the decomposition method which describes how to separate light from the hot spot and the disk. This tool was applied successfully for such eclipsing CVs as Z Cha, OY Car (Smak 2007, 2008ab), HBHA 4705-03 (Rutkowski et al. 2013) and HT Cas (B\k{a}kowska \& Olech 2014).

After the decomposition procedure for eclipses presented in Fig.3, we obtained four separate light curves of the hot spot and the disk. The uneclipsed parts of the resulting hot spot light curves were fitted with a formula given by Paczy\'{n}ski \& Schwarzenberg-Czerny (1980)
\begin{equation}
l^*_{s,o}(\phi)=A_{s,max}[1-u+u\cos(\phi-\phi_{max})]\cos(\phi-\phi_{max})
\end{equation} 
where the limb darkening coefficient was $u=0.6$. In Table 2 we present the hot spot amplitude $A_{s,max}$, phase of its maximum $\phi_{max}$ and four successive phases of contacts $\phi_1$, $\phi_2$, $\phi_3$ and $\phi_4$. 

\begin{table*}
\begin{footnotesize}
 \begin{center}
  \caption{Hot Spot in IY UMa. Hump parameters and phase contacts.}
  \begin{tabular}{@{}lllllll@{}}
  \hline
   Date of     &  Phase of hot spot   & Hot spot  & Phases of &  &  & \\
  eclipse  					 &  max. brightness      &         amplitude           & contact  & & & \\
   & $\phi_{max}$ & $A_{s,max}$[\%]& $\phi_1$ & $\phi_2$ & $\phi_3$ & $\phi_4$ \\
\hline
2012 Apr 17/18 & $-0.070\pm0.003$ & $47.31\pm0.61$ & $-0.065\pm0.01$ & $-0.012\pm0.005$ & $0.075\pm0.005$ & $0.105\pm0.015$\\ 
2013 Mar 05/06 & $-0.064\pm0.005$ & $24.15\pm0.98$ & $-0.056\pm0.005$ & $-0.034\pm0.003$ & $0.060\pm0.005$ & $0.087\pm0.01$\\ 
2013 Mar 06/07 & $-0.047\pm0.002$ & $47.24\pm0.03$ & $-0.060\pm0.005$ & $-0.032\pm0.003$ & $0.065\pm0.005$ & $0.095\pm0.01$\\ 
2015 Oct 14/15 & $-0.056\pm0.007$ & $42.65\pm0.54$ & $-0.049\pm0.002$ & $-0.037\pm0.002$ & $0.060\pm0.005$ & $0.078\pm0.005$\\ 
\hline
\end{tabular}
\label{tab3}
\end{center}
\end{footnotesize}
\end{table*}

\vspace*{18.50cm}
\includegraphics{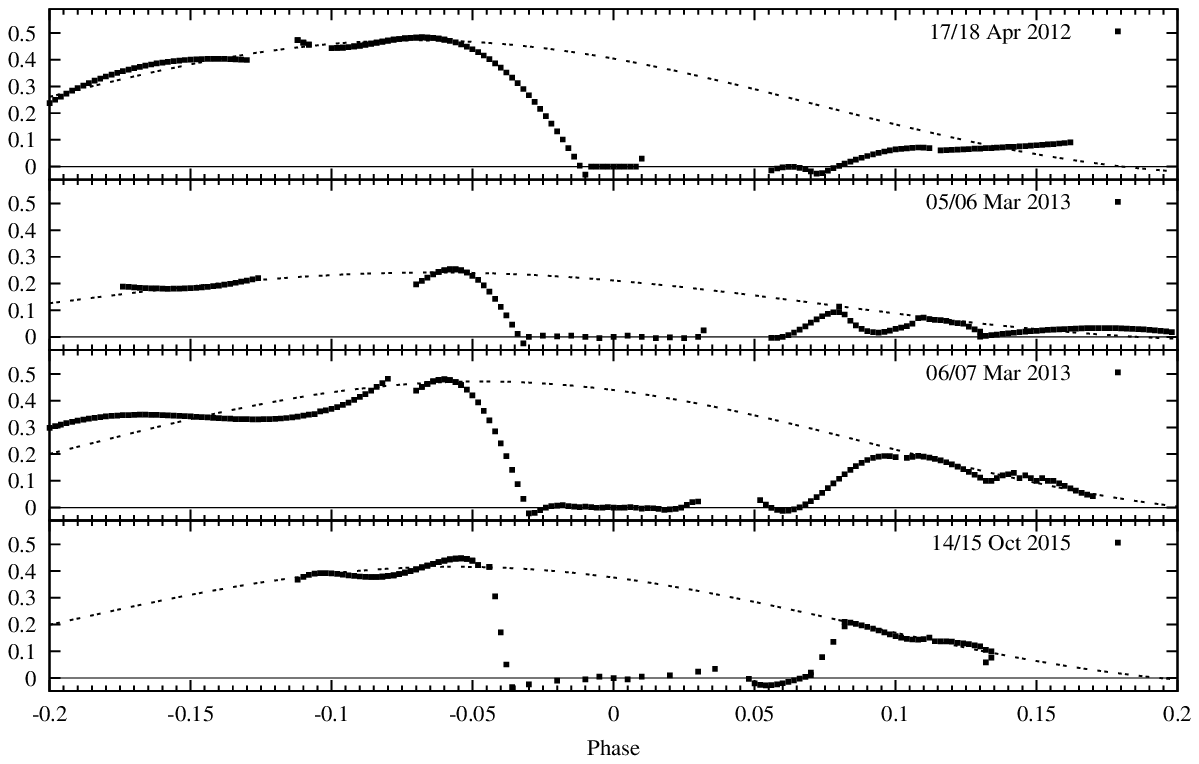}
\vspace*{-4.0cm}
   \begin{figure}[!ht]
      \caption {Reconstructed hot spot light curves of IY UMa. Dotted lines represent Eq.(1) with parameters listed in Table 2. The hot spot manifestation with the brightness amplitude over $40\%$ is clearly seen in the panels 1, 3 and 4 (from the top), so before and after normal outburst. During the normal outburst (second panel from the top), the brightness amplitude of the hot spot dropped to the level below $25\%$.}
   \end{figure}

In Fig.4 (from the top panel to the bottom) the spot light curves were displayed. The amplitude of brightness of the hot spot was enormous (reaching over $40\%$) before the outburst (top panel).  In the second panel (from the top) the reconstructed hot spot from the March 2012 outburst is presented and one can see that the hump amplitude dropped dramatically below $25\%$. In the third and the fourth panels (from the top) we present the hot spot after the outburst, the hump amplitude was again on its average level (above $40\%$).

In Fig.5 we display the light curve of the IY UMa during quiescence and the normal outburst (top panel) and the amplitude of the hot spot (bottom panel). In our opinion, there is a strong dependence between the activity of IY UMa and the hot spot brightness.

\vspace*{14.50cm}
\includegraphics{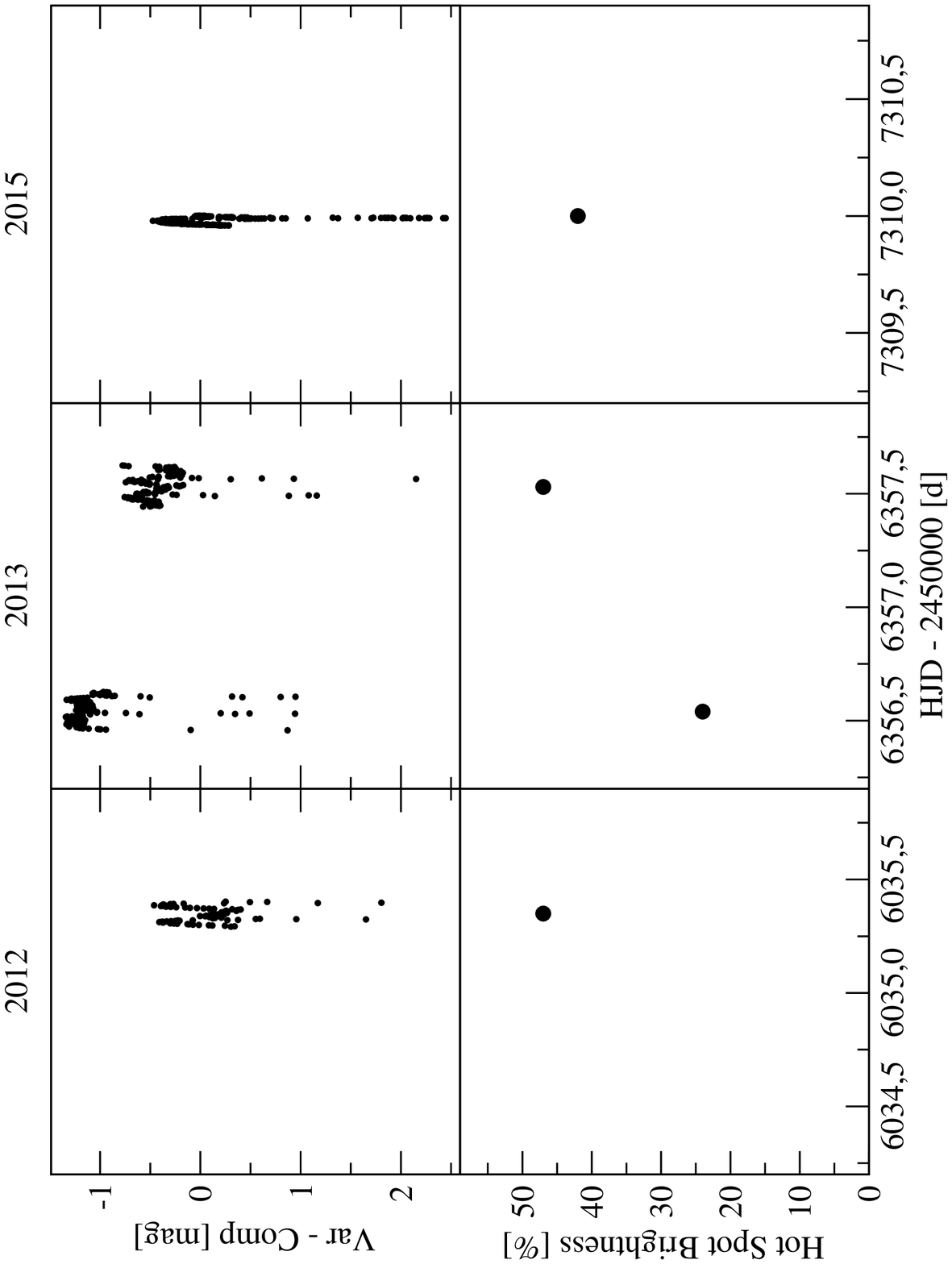}

   \begin{figure}[!ht]
      \caption {In the top panel we present light curves of IY UMa used for the decomposition analysis. Our data set covers quiescence before (17/18 April 2012, left top panel) and after (14/15 October 2015, right top panel) the normal outburst of IY UMa. The middle top panel displays IY UMa during the last night of the normal outburst and during the quiescence on the successive night (05/06 and 06/07 March 2013, respectively). In the bottom panel we display the amplitude of the hot spot. During quiescence the hot spot amplitude was above $40\%$.  Throughout the normal outburst, the hot spot amplitude dropped to the level below $25\%$.}
   \end{figure}

\section{Discussion}

Lately, there has been an intense discussion about DNs models (for review B\k{a}kowska \& Olech 2014). Analysis of eclipsing DNs (Smak 1994bc, 2007, 2008ab, B\k{a}kowska \& Olech 2014) turned out to be milestones in favour of the EMT model in a recent debate about the superoutbursts and superhump mechanisms. 

During normal outbursts in DNs the accretion disk brightens by factor 10-100 for a few days as a result of a sudden increase in mass accretion. Bath (1975) proposed the mass transfer instability model (MTI) where the outbursts are caused by a burst of mass transfer from the donor star. In the thermal-viscous disk-instability model (DIM) a bimodal behaviour of the viscosity as a function of the surface density is responsible for the sudden increase in the mass transfer through the disk (see reviews by Smak 1984, Lasota 2001).The behaviour of the bright spot of IY UMa during its March 2012 outburst is similar to the one observed in OY Car (check Fig.1 in Rutten et al. 1992). The April 1979 normal outburst in OY Car was observed by Voght (1981, 1983) and reanalized by Rutten et al. (1992) and before the normal outburst the spot was clearly visible but it was gradually disappearing during the development of the outburst which offered no support for the mass-transfer instability model. However, Baptista (2012) analysed the mechanisms of accretion disks' evolutions during outbursts in several eclipsing DNs. Based on the observational evidences there are two classes of DNs and during the outbursts one group follows the DIM scenario (e.g. Z Cha, OY Car, IP Peg) and while the behaviour of the others  (e.g. EX Dra, HT Cas, V2051 Oph) can be explained in terms of the MTI. In case of our data set of IY UMa observations during normal outburst, we conclude that IY UMa behaves in accordance with the DIM framework. Nevertheless, the models describing normal outbursts should continue to be analysed since no consensus has been reached yet.

The decomposition method for the eclipsing DNs to investigate the hot spot brightness was applied only a few times (check Table 3 in B\k{a}kowska \& Olech 2014) but it has already proven useful as it shed some light on the superoutburst mechanism. In the case of HT Cas time span between two successive superoutbursts is about 25 years, also normal outbursts are a unique and rare phenomenon. According to Kato et al. (2009) the superoutbursts in IY UMa were detected in 2000, 2002, 2004, 2006 and 2009. Nonetheless, the analysis of the bright spot of IY UMa before, during and after the active state has not been made yet. Observations of normal outbursts in eclipsing DNS are also of great value, i.e. Smak (1995) determined mass transfer rates in Z Cha and U Gem based on extensive photometric coverage of the luminosity of the hot spot. Rolfe et al. (2005) showed in IY UMa the dramatic changes in the accretion flow and emission at the onset of an outburst. Unfortunately, our data does not cover the beginning of the March 2012 outburst, also we did not collect any observations for a few successive normal outbursts. It is also worth noting the observations of IY UMa during quiescence made by Steeghs et al. (2003) and reanalized by Smak (2003) who noticed that about $10\%$ of the observed disk flux comes from parts of the stream which are overflowing the disk. Stanishev et al. (2000) conducted the eclipse mapping of IY UMa during the late decline of the 2000 superoutburst and in quiescence stages and the strength of the orbital hump decreased by $\sim30\%$ which further supports the idea that such a decrease could be produced by an enhanced mass transfer rate.

Lately, there have been huge surveys dedicated for CVs (Kato et al. 2009, 2010, 2012, 2013, 2014ab) or the other projects where CVs are detected and observed when an opportunity arises, such as ASAS-SN (Davis et al. 2015) or OGLE surveys (Soszy\'{n}ski et al. 2015, Mr\'{o}z et al. 2013, Poleski et al. 2011). Eclipsing DNs can be investigated with tools such as the eclipse mapping method (Horne 1985) or the decomposition method (Smak 1994a) and that is why these objects are of great value of the future development of the theory of CVs.

\section{Conclusions}

We present the analysis of the hot spot brightness in eclipsing dwarf nova IY UMa before, during and after its normal outburst. In order to obtain separated light curves from the hot spot and from the disk, we used the decomposition method invented by Smak (1994a). Based on our data, the hot spot amplitude changed significantly during the normal outburst. During quiescence, before and after the outburst, the average bright spot amplitude was above $A>40\%$. During the March 2013 outburst in IY UMa, the brightness of the hot spot was below $A<25\%$. In our opinion, the behaviour of the bright spot shows that IY UMa during its normal outburst follows the DIM model.

After intense observations of IY UMa at the beginning of this century, this eclipsing DN has received very little attention during last decade. IY UMa is definitely worth observers' attention. This star is an excellent accretion laboratory because of its high inclination, frequent outbursts and superoutbursts and brightness available for 1-meter class telescopes. 

\bigskip

\noindent {\bf Acknowledgments.} ~We acknowledge the generous allocation of the Warsaw Observatory 0.6-m  telescope time. Moreover, this work is based on observations obtained at the MDM Observatory, operated by Dartmouth College, Columbia University, Ohio State University, Ohio University, and the University of Michigan. KB wants to thank to K.Z. Stanek, J.R. Thorstensen, E. Alper and E. Galayda for all the technical support and tips about observations in MDM Observatory. We also want to thank mgr A. Borowska for language corrections and R.Pospieszy\'{n}ski for his help in reduction and photometry analysis. The project was supported by the Polish National Science Center grants awarded by decisions DEC-2012/07/N/ST9/04172 and DEC-2015/16/T/ST9/00174 for KB.


\begin{thebibliography}{}

\bibitem{Baptista12} Baptista R., 2012, {\it Memorie della Societa Astronomica Italiana}, 83, 530

\bibitem{Bath75} Bath G.T., 1975, {\it MNRAS}, 171, 311

\bibitem{Bak2014} B\k{a}kowska K. \& Olech A., 2014, {\it Acta Astron.}, 64, 247

\bibitem{Davis2015} Davis A.B., Shappee B.J. \& Archer Shappee B., 2015, {\it American Astronomical Society Meeting Abstracts}, 225, 344

\bibitem{Hell00} Hellier C., 2001,  {\it Cataclysmic Variable Stars}, Springer.

\bibitem{Horne85} Horne K., 1985, {\it MNRAS}, 213, 129

\bibitem{Rolfe05} Rolfe D.J., Haswell C.A., Abbott T.M.C., Morales-Rueda L., Marsh T.R. \& Holdaway. 2005, {\it MNRAS}, 357

\bibitem{Kato01} Kato T., Stubbings R., Nelson P., Pearce A., Garradd G. \& Kiyota S.. 2001, {\it IBVS}, 5159, 0374

\bibitem{Kato09} Kato T. et al., 2009, {\it PASJ}, 61, S395

\bibitem{Kato10} Kato T. et al., 2010, {\it PASJ}, 62, 1525

\bibitem{Kato12} Kato T. et al., 2012, {\it PASJ}, 64, 21

\bibitem{Kato13} Kato T. et al., 2013, {\it PASJ}, 65, 23

\bibitem{Kato14a} Kato T. et al., 2014a, {\it PASJ}, 66, 30

\bibitem{Kato14b} Kato T. et al., 2014b, {\it PASJ}, 66, 90

\bibitem{Lasota01} Lasota J.-P., 2001, {\it New Astronomy Reviews}, 45, 449

\bibitem{Mroz13} Mr\'{o}z P. et al., 2013, {\it Acta Astron.}, 63, 135

\bibitem{Paczynski80} Paczy\'{n}ski B. \& Schwarzenberg-Czerny A., 1980, {\it Acta Astron.}, 30, 127

\bibitem{Pat00} Patterson J. et al., 2000, {\it PASP}, 112, 1567

\bibitem{Poleski11} Poleski R. et al., 2011, {\it Acta Astron.}, 61, 123

\bibitem{Rolfe05} Rolfe D.J., Haswell  C.A., AbbottT.M.C., Morales-Rueda L., Marsh T.R. \& Holdaway G., 2005, {\it MNRAS}, 357, 69


\bibitem{Rutkowski13} Rutkowski A., Ak T., Marsh T.R. \& Eker Z., 2013, {\it Acta Astron.}, 63, 225

\bibitem{Rutten92} Rutten R.G.M., Kuulkers E., Vogt N. \& van Paradijs J. 1992, {\it A\&A}, 265, 159

\bibitem{Soszynski2015} Soszy\'{n}ski I. et al, 2015, {\it Acta Astron.}, 65, 39

\bibitem{Smak84} Smak J., 1984, {\it PASP}, 96, 5

\bibitem{Smak94a} Smak J., 1994a, {\it Acta Astron.}, 44, 45

\bibitem{Smak94b} Smak J., 1994b, {\it Acta Astron.}, 44, 59

\bibitem{Smak94c} Smak J., 1994c, {\it Acta Astron.}, 44, 256

\bibitem{Smak95} Smak J., 1995, {\it Acta Astron.}, 45, 355

\bibitem{Smak03} Smak J., 2003, {\it Acta Astron.}, 53, 167

\bibitem{Smak07} Smak J., 2007, {\it Acta Astron.}, 57, 87

\bibitem{Smak08a} Smak J., 2008a,{\it  Acta Astron.}, 58, 55

\bibitem{Smak08b} Smak J., 2008b,{\it  Acta Astron.}, 58, 65

\bibitem{Stanishev01} Stanishev V., Kraicheva Z., Boffin H.M.J. \& Genkov V., 2001, {\it A\&A},  367, 273

\bibitem{Steeghs03} Steeghs D. et al., 2003, {\it MNRAS}, 339, 810

\bibitem{Stetson87} Stetson P.B., 1987, {\it PASP}, 99, 191

\bibitem{Tak98} Taamizawa K., 1998, vsnet-obs circulation 18078 (http://www.kusastro.kyoto-u.ac.jp/vsnet/Mail/obs18000/msg00078.html)

\bibitem{Udalski92} Udalski A. \& Pych W., 1992, {\it Acta Astron.}, 42, 285

\bibitem{Ue00} Uemura M. et al., 2000, {\it PASJ}, 52, L9

\bibitem{Vogt81} Vogt N., Schoembs R., Krzeminski W. \& Pedersen H., 1981,{\it  A\&A }, 94, L29

\bibitem{Vogt83} Vogt N., 1983, {\it  A\&A}, 128, 29

\bibitem{War00} Warner B., 1995,  {\it Cataclysmic Variable Stars}, Cambridge University Press.

\end{thebibliography}
\end{document}